\documentstyle[12pt]{article}  
\begin{document}
\newcommand{\be}{\begin{equation}}
\newcommand{\ee}{\end{equation}}
\newcommand{\ben}{\begin{eqnarray}}
\newcommand{\een}{\end{eqnarray}}
\vskip 0.4in
\centerline{\Large\bf Incompatibility of the de Broglie-Bohm}
\vskip 0.1in
\centerline{\Large\bf Theory with Quantum Mechanics}
\vskip 0.4in
\centerline{\large Partha Ghose}
\vskip 0.1in
\centerline{S. N. Bose National Centre or Basic Sciences}
\vskip 0.1in
\centerline{Block JD, Sector III, Salt Lake, Calcutta 700 091, India}
\vskip 0.2in
\begin{abstract} It is shown that although the de Broglie-Bohm quantum theory of motion is equivalent to standard quantum mechanics when averages of dynamical variables are taken over a Gibbs ensemble of Bohmian trajectories, the equivalence breaks down for ensembles built over clearly separated short intervals of time in special multi-particle systems. This feature is exploited to propose a realistic experiment to distinguish between the two theories.
\end{abstract}
\vskip 0.4in
PACS No. 03.65.Bz
\vskip 0.4in
The proponents of the de Broglie-Bohm quantum theory of motion (dBB)
\cite{Bohm}, \cite{Bohm2}, \cite{Holland} have always held that it
is constructed to make exactly the same
{\it statistical} predictions as standard quantum theory (SQT) in every
conceivable physical situation. By statistical predictions is meant averages
of dynamical variables over all possible ``hidden'' trajectories at a fixed instant of time (a virtual Gibbs ensemble). The average over a Gibbs ensemble usually turns out to be the same as the average over a time ensemble, i.e., an ensemble built over clearly separated short intervals of time. Although this equivalence holds in SQT, I will show that it breaks down for two-particle systems in dBB when a certain combination of bosonic and geometrical symmetries holds.
It is possible to exploit this feature and do an experiment with a sequence of photon pairs that can distinguish between SQT and dBB, so far thought to be statistically completely equivalent.

When Bohm proposed his theory in 1952, only experiments with beams of
particles (Gibbs ensembles) were possible, and he constructed his theory 
in such a fashion
as to make it impossible to discriminate between it and SQT using such
experiments. But by its very {\it raison de \^ etre} the theory is
essentially different from SQT at the level of individual systems which are
precisely defined and deterministic. Indeed, he wrote in the
abstract of his first paper on hidden variables \cite{Bohm},
\begin{quotation}
`The usual interpretation of the quantum theory is
self-cocnsistent, but it involves an assumption that cannot be tested
experimentally, {\it viz}, that the most complete possible specification
of an individual system is in terms of a wave function that determines only
probable results of actual measurement processes. The only way of
investigating the truth of this assumption is by trying to find some other
interpretation of the quantum theory in terms of {\it at pressent} ``hidden''
variables, which in principle determine the precise behaviour of an
individual system, but which are {\it in practice averaged over} in
measurements of the type that can {\it now} be carried out. ... the
suggested interpretation provides a broader conceptual framework than the
usual interpretation, because it makes possible a precise and continuous
description of all processes, even at the quantum level.' (Italics mine)
\end{quotation}
The possibility that the ``hidden'' variables (the trajectories of the particles in dBB) can become ``visible'' (if not directly, at least through their unambiguous signals) in experiments of a different type (involving a time sequence of single systems, which can nowadays be carried out) is clearly left open by this statement.

Let us closely examine the usual argument establishing the equivalence of the two theories. Consider $n$ identical particles of mass $m$ and write their normalized wave function (symmetrized or antisymmetrized in the particle coordinates) in the polar form

\be
\psi(x_1,x_2,...,x_n,t) = R(x_1,x_2,...,x_n,t) e^{\frac{i}{\hbar} S(x_1,x_2,...,x_n,t)}
\ee
where $x_i$ stand for the three dimensional vectors $\vec{x}_i$. In addition to the wave function one also introduces the positions of the particles $(x_1,x_2,...,x_n)$ in dBB by first defining the velocities of the particles through the first-order differential equations
\newpage
\ben
v_i(x_1,x_2,...,x_n,t)\equiv \frac{d x_i}{d t} &=& \frac{1}{m} \partial_{x_i} S(x_1,x_2,...,x_n,t)\nonumber\\
&=& \frac{\hbar}{m} Im \frac{\partial_{x_i}\psi(x_1,x_2,...,x_n,t)}{\psi(x_1,x_2,...,x_n,t)}
\label{eq:0}
\een
and integrating them. Varying the initial points in $3n$ dimensional configuration space, one obtains the set of {\it deterministic} Bohmian trajectories of the system which describes the behaviour of every particle in the system. These trajectories are given an ontological status in dBB.

Since the representative points in configuration space vary randomly, use is made of the identity $\vert \psi(x_1,x_2,...,x_n,t)\vert^2 = R^2(x_1,x_2,...,x_n,t)$ to define their density function as

\be
P (x_1(t),x_2(t),...,x_n(t),t)= R^2(x_1,x_2,...,x_n,t)
\label{eq:01}
\ee
The reason for this choice is that the quantum mechanical probability density $R^2$ is conserved,

\be
\frac{\partial R^2}{\partial t} + \sum_{i}\partial_{x_i} (R^2 \partial_{x_i} S/m) = 0
\label{eq:02}
\ee
as a consequence of the Schr\"{o}dinger equation,
and hence the identification (\ref{eq:01}) guarantees that the density function $P$ in dBB is also conserved:

\be
\frac{\partial P}{\partial t} + \sum_{i}\partial_{x_i}( P v_i) = 0
\label{eq:03}
\ee
Thus, once the initial density function $P(t=0)$ is taken to be the same as implied by the quantum mechanical expression $R^2(0)$, the continuity equations (\ref{eq:02}) and (\ref{eq:03}) guarantee that the averages of all dynamical variables $O$ of the system computed over the measure $P$ (in the same sense as in classical statistical mechanics) will necessarily agree with the quantum mechanical expectation values of the corresponding hermitian operators $\hat{O}$ at all future times $t$ \cite{Holland}:

\be
\langle O\rangle = \int_{t=constant} P O d^{3n} x = \int_{t=constant} \psi^* \hat{O} \psi d^{3n} x
\ee
This completes the proof of equivalence.

The crucial point to notice in this proof is the following. It follows from the definition of the Bohmian velocities (\ref{eq:0}) that the convection currents $R^2\partial_{x_i}S/m$ in the continuity equation (\ref{eq:02}) are replaced by $P v_i$ in the continuity equation (\ref{eq:03}). This implies that the Bohmian trajectories in $3n$ dimensional configuration space do not touch or cross one another and constitute a phase flow because of the uniqueness of the gradients of the action function $S$. The mapping of the initial density function $P(t=0)$
in configuration space to the final density function $P(t)$ in the target space therefore not only conserves the density, {\it it also contains information about the causal histories of the particles which are  distinguishable}. The corresponding quantum mechanical mapping does not have this information, and the particles remain indistinguishable and without any consistent histories. There is therefore an essential difference between these two mappings which leads not only to physical predictions in dBB that cannot be made unambiguously in SQT (for example, tunneling times \cite{Leavens}) but also to predictions that are {\it inconsistent} with SQT in at least one special case, as I will show in what follows.

One might wonder how the distinguishability of the particles is reconciled
with their bosonic or fermionic behaviour in dBB. The purely quantum
correlations between the particles are, in fact, brought about by the
quantum potential

\be
Q(x_1,x_2,...,x_n) = - \frac{\hbar^2}{2 m R}\sum_{i=1}^{n} \nabla_{x_i}^2 R(x_1,x_2,...,x_n)
\ee
acting between them. Einstein was already aware of this
problem in 1925 when he referred to the `mysterious interactions' that
bring about the difference between Maxwell-Boltzmann and
Bose-Einstein statistics \cite{Pais}. As is well known, neither Bose nor
Einstein introduced the concept
of {\it indistinguishability} in their seminal papers in 1924/25 on quantum
statistics. The concept of indistinguishability entered into quantum physics only after Schr\"{o}dinger introduced the wave function and the particle
position was jettisoned in favour of it.

To see how predictions of dBB and SQT can defer in a special case, let us  consider a variant of the double-slit experiment
in which the wave packets of a pair of identical (momentum
correlated) non-relativistic particles $1$ and $2$ are simultaneously 
diffracted by two slits $A$ and $B$ (which can be regarded as the sources of 
the diffracted waves) and overlap in some region $\cal{R}$ (see figure)
sufficiently 
far from the slits so that the travelling waves in that region can be 
regarded as plane waves.
Let us assume that the particles are bosons. The problem is essentially a
two dimensional one. So let this plane be the $x-y$ plane.
Then the two-particle wave function in the region $\cal{R}$ in the $x-y$ plane can be written as (\cite{Holland})

\ben
\Psi \biggl((x_1, y_1), (x_2, y_2), t\biggr) = \langle (x_1, y_1), (x_2, y_2), t
\frac{1}{\sqrt{2}} \left[ \vert 1_{A}, 2_{B} \rangle + \vert 1_{B}, 2_{A}
 \rangle \right]\nonumber
\een
\ben
= \frac{1}{\sqrt{2}} [ \psi_{A}( x_1, y_1, t ) \psi_{B} (x_2, y_2, t ) +  \psi_{A} (x_2, y_2, t ) \psi_{B} (x_1, y_1, t ) ]
\label{eq:1}
\een
Notice that the particles lose their identities and become
indistinguishable in SQT as a result of this symmetrization.
The probability of joint detection
of the particles around points $x(P)$ and $x(Q)$ on a screen with fixed $y$
in the region $\cal{R}$ is given by

\be
P_{1 2} \biggl(x (P), x(Q)\biggr) \delta x(P) \delta x(Q) =
\int_{x(P)}^{x(P) + \Delta x (P)} d x_1
\int_{x(Q)}^{x(Q) + \Delta x(Q)} d x_2 \vert \Psi \vert^{2}
\label{eq:2}
\ee
which is to be evaluated on a $t=$ constant spatial surface. It contains fourth-order interference terms between $1$ and $2$
\cite{Ghosh}, \cite{Horne}. (The small but finite domains of integration are chosen to take account of the finite size of the detectors.)

Now, consider the $x$-components
of the Bohmian velocities of particles $1$ and $2$,

\ben
v_{x}(1)&=& \frac{\hbar}{m} Im \frac{\partial_{x_{1}}
\Psi \biggl( (x_1, y_1), (x_2, y_2), t\biggr)}{\Psi \biggl((x_1, y_1), (x_2, y_2), t\biggr)} |_{(x_{1},y_{1}) = ( x_{1},y_{1})(t)}\nonumber\\
v_{x}(2) &=& \frac{\hbar}{m} Im \frac{\partial_{x_{2}}
\Psi \biggl( (x_1, y_1), (x_2, y_2), t\biggr)}{\Psi \biggl((x_1, y_1), (x_2, y_2), t\biggr)} |_{(x_{2},y_{2}) = (x_{2},y_{2})(t)}
\label{eq:5}
\een
Although the configuration space trajectories of the $2$-particle system do not touch or cross one another, their projections on to $3$ dimensional coordinate space (i.e., the individual trajectories of the two particles in ordinary space) can cross in general. However, the double-slit arrangement has a natural symmetry line that bisects the line joining the two slits. Because of this geometrical symmetry of the arrangement, we have

\ben
\psi_{A} (x_1, y_1, t) &=& \psi_{B} (-x_1, y_1, t)\nonumber\\
\psi_{A} (x_2, y_2, t) &=& \psi_{B} (-x_2, y_2, t)
\label{eq:6}
\een
which implies that the wave function given by eqn. (\ref{eq:1}) is symmetric under reflection in the plane $x = 0$. It follows from this and equation (\ref{eq:5}) that

\ben
v_{x}(1)(x_1,x_2, y_1,y_2,t) &=& - v_{x}(1)(-x_1,-x_2, y_1,y_2,t)\\
v_{x}(2)(x_1,x_2, y_1,y_2,t) &=& - v_{x}(2)(-x_1,-x_2, y_1,y_2,t)
\label{eq:7}
\een
This implies that the $x$
components of the velocities of the two particles would vanish 
on the plane of symmetry provided both of them were simultaneously on this
plane, i.e., $x_1(t) = x_2(t) = 0$. {\it In that event they cannot touch or 
cross this plane}.

This is guaranteed by translation invariance in the region $\cal{R}$ where the wave function
$\Psi$ (eqn. (8)) can be written as the product $\Phi(\vec{X}) \phi(\vec{x})$ 
where
$\vec{X}=\frac{1}{2} (\vec{x}_1 + \vec{x}_2)$ is the centre-of-mass 
coordinate and $\vec{x}=
(\vec{x}_1 - \vec{x}_2)$. It follows from this that the only non-trivial
phase is that of the wave function 
$\phi(\vec{x})$ which can be a function only of
$(\vec{x}_1 - \vec{x}_2)$. Since the overall uniform motion 
of the centre-of-mass
of the system is along the $Y$ direction because of the symmetry of the set-up, 
it follows from the definition of Bohmian velocities (\ref{eq:0}) 
that $\dot{x_1} + \dot{x_2} = 0$ 
which implies

\be
x_1(t) + x_2(t) = x_1(0) + x_2(0)
\label{eq:8}
\ee
Therefore, if $x_1(0) + x_2(0) = 0$, it follows that the  
trajectories of the two 
particles will always be symmetrical about the plane of symmetry $x=0$.

It is important to draw attention to the crucial role played by the combination of bosonic and geometric symmetry in the above argument. Notice that if bosonic symmetry is not imposed on the two-particle wave function (equation (\ref{eq:1})), it will not be symmetric under reflection in the plane $x=0$. This will mean that with Maxwell-Boltzmann statistics (only one term in equation (\ref{eq:1})), for example, the particles $1$ and $2$ can cross one another and the plane $x=0$. But the situation is different with bosonic symmetry. This additional symmetry constrains even the individual particle trajectories in coordinate space not to touch or cross one another and this plane \cite{Holland}. This brings about a fundamental change---the  coordinate space particle trajectories are clearly separated into two non-intersecting disjoint classes, one above and one below the plane of symmetry. This is the source of the basic incompatibility between dBB and SQT.

To see what this implies for the two-particle double-slit experiment under consideration, consider the dBB ensemble to be built up of single pairs of particle trajectories arriving at the screen at different instants of time $t_i$ such that the joint probability of detection is given by

\ben
P_{1 2}
= lim_{N \rightarrow \infty}\sum_{i=1}^N
\frac{1}{\delta (0)}\int d x_1 \int d x_2 P(x_1,x_2,t_i)\nonumber\\
\delta (x_1 - x_1(t_i))\,\delta (x_2 - x_2(t_i)) \delta (x_1(t_i) + x_2(t_i))\nonumber\\
= lim_{N \rightarrow \infty} \sum_{i=1}^N P(x_1(t_i), -x_1(t_i)) = 1
\label{eq:9}
\een
where the constraint (\ref{eq:8}) has been taken into account. Every term in the sum represents only one pair of trajectories arriving at the screen at the points $(x_1(t_i), -x_1(t_i))$ at time $t_i$, weighted by the corresponding density $P$. In discrete time ensembles of this type every pair can be separately identified, and it is clear that if the detectors are placed symmetrically about the plane $x=0$, they will record coincidence counts just as predicted by SQT. On the other hand, if they are placed asymmetrically about $x=0$, the joint detection of every pair, and hence also their time average, will produce a null result which is {\it in conflict with the SQT prediction} \cite{footnote}.

If all the times $t_i$ are put equal to a fixed time $t$ in (\ref{eq:9}), the sum over $i$ can be converted to an integral over all the trajectories which pass through all the points of the screen at that time (a Gibbs ensemble). Now consider the joint probability of detection around two points $x(P)$ and $x(Q)$ on the screen that are not symmetric about $x=0$. The trajectories that pass through these regions cannot be partners of individual pairs which are constrained by (\ref{eq:8}), and must therefore belong to {\it different} pairs which are {\it not so constrained}. If the points are symmetrically situated about $x=0$, the constraint is automatically satisfied. The joint probability of detection in this case is therefore given by
\ben
P_{12}\biggl(x(P),x(Q),t\biggr)
= \int_{x(P)}^{x(P) + \Delta x (P)} d x_1(t)
\int_{x(Q)}^{x(Q) + \Delta x(Q)} d x_2(t)\nonumber\\
P(x_1(t), x_2(t))
\een
which is, in fact, the same as the SQT prediction (\ref{eq:2}) by virtue of the dBB postulate (\ref{eq:01}). It becomes impossible therefore to distinguishing between dBB and SQT using such an experiment, as correctly maintained by Bohm.

Nevertheless, the two theories actually deal with entirely different types of fundamental entities. The particles in dBB are {\it in principle distinguishable and deterministic} and have definite histories that remain ``hidden'' in experiments with beams of particles, whereas they are {\it in principle indistinguishable and indeterministic} and have no histories at all in SQT. Thus, the statistical distribution of the particles are the same in the two theories only to the extent that one can ignore their histories and distinguishability. Once distinguishability through the trajectories is admitted as in dBB, it leads to an intrinsic difference between the ensembles in the two theories which can only be resolved by using time ensembles. It was not possible to produce such ensembles in 1952 when Bohm proposed his theory, but in recent times it has become possible to do so by using, for example, two-photon radiative cascades or parametric down-conversion of laser pulses into ph!
oton pairs and cutting down their 
intensities to single particle levels using neutral density filters.

Although a similar separation of the trajectories into two
disjoint non-intersecting classes above and below the plane of symmetry also occurs in the standard double-slit experiment with single particles, it does {\it not} lead to different resuts in the two theories, because with a single particle the question of ditinguishability does not arise. Both theories predict anti-coincidences in this case for all possible positions of two counters, simply because only one particle passes through the apparatus at
any given time.

Exactly the same conclusion can be drawn for a pair of photons
by using a consistent quantum mechanical formalism for massless
relativistic bosons below the threshold of pair creation \cite{Ghose}. In an actual experiment with photons,
one must choose identical photons generated by degenerate parametric
down-conversion, for example, and make the signal and idler photons pass
through the two slits $A$ and $B$ simultaneously. One must also cut down 
the intensities of the beams to the single pair level. To the best of 
my knowledge these conditions have not been met with in any experiment 
done so far \cite{Ghosh}. Hence the necessity for a critical experiment 
which can settle the fundamental questions as to whether the wave function 
description is the most complete possible and also whether the lack of 
causality in quantum phenomena is really fundamental.

I am grateful to Franco Selleri, Rupa Ghosh,
S. M. Roy, C. S. Unnikrishnan, A. S. Majumdar and B. Dutta-Roy for many
helpful
discussions. I also wish to acknowledge financial support from the 
Department of Science and Technology, Government of India, through a 
research grant.

\vskip 0.2in

FIGURE CAPTION: Two-particle double-slit experiment

\pagebreak

\setlength{\unitlength}{1mm}
\begin{picture}(120,100)
\put(1,0){\line(1,0){105}}
\put(105,0){\vector(1,0){18}}
\put(10,0){\line(0,1){15}}
\put(10,0){\line(0,-1){15}}
\put(10,-12){\line(0,-1){10}}
\put(10,12){\line(0,1){10}}
\put(10,25){\line(1,1){70}}
\put(10,24){\line(1,-1){70}}
\put(10,-24){\line(1,1){70}}
\put(10,-25){\line(1,-1){70}}
\put(10,27){\vector(0,1){18}}
\put(10,-27){\vector(0,-1){18}}
\put(80,0){\line(0,1){100}}
\put(80,0){\line(0,-1){100}}
\put(67,4){\LARGE $\cal{R}$}
\put(5,45){x}
\put(7,1){0}
\put(120,-3){y}
\put(5,23){A}
\put(5,-25){B}
\put(83,10){P}
\put(83,-23){Q}
\put(80,10){-}
\put(80,-18){-}

\end{picture}


\begin{thebibliography}{99}
\bibitem{footnote}
Implicit in this is the necessary requirement in dBB that only particles produce detections and not `empty' parts of wave functions. See Ref. 3.
\end{thebibliography}
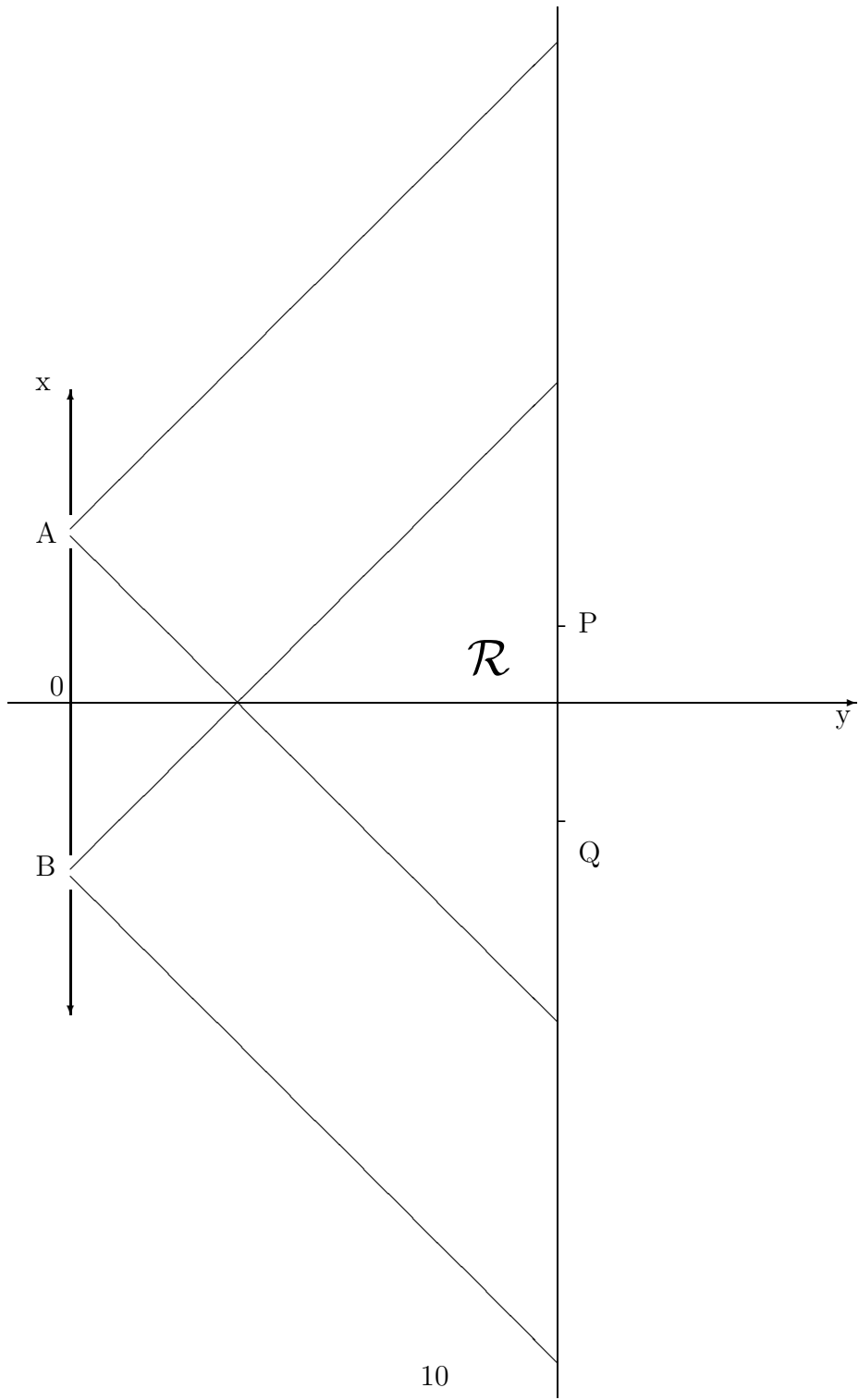
\end{document}